\documentstyle[aps,preprint,epsfig]{revtex}

\begin{document}

\bibliographystyle{unsrt}
\title{Observation of Nonlinear Mode in a Cylindrical Fabry-Perot Cavity}
\author{Jack Boyce, Juan P. Torres and Raymond Y. Chiao}
\address{
Department of Physics, University of California, Berkeley, 
California 94720 \\
Voice: (510) 642-5620 Fax: (510) 643-8497}
\maketitle

\begin{abstract}
We report the first observation of a nonlinear mode in a 
cylindrical nonlinear Fabry-Perot cavity.  The field enhancement from 
cavity buildup, as well as the large $\chi^{(3)}$ optical nonlinearity 
due to resonantly-excited $^{85}\mathrm Rb$ vapor, allows the 
nonlinear mode to form at low incident optical powers of less than a 
milliwatt.  The mode is observed to occur for both the self-focusing 
and self-defocusing nonlinearity.
\end{abstract}

\newpage

Self-trapping and self-focusing effects due to the nonlinear response
of materials to an applied optical field have been an interesting
topic of research from the early days of nonlinear optics
\cite{chiao1964}.  Most work has been done in traveling-wave systems,
where waveguides and pulsed lasers are used to achieve large
intensities and enhanced nonlinear effects.  Spatial solitons have
become a widespread paradigm in explaining many nonlinear
effects~\cite{akhmediev}, and have been observed in
Kerr~\cite{barthelemy1985,aitchison1990},
$\chi^{(2)}$~\cite{schiek,torruellas}, and
photorefractive~\cite{segev} media.  In a cavity geometry, the main
concern has been to explore optical bistability for its potential
applications such as all-optical switching~\cite{gibbs}.

In a linear cavity there are specific beam profiles that result in
transmissivity, defined as the ratio of the transmitted power to the
incident power, being equal to unity.  These are the so-called
longitudinal and transverse modes of the cavity.  In a cavity with
planar mirrors these modes are plane waves; such a cavity driven by a
finite-sized beam yields a transmissivity less than unity due to
diffraction.  Notwithstanding, peaks in the transmissivity exist for
frequencies corresponding to the longitudinal modes.  When the cavity
is filled with a nonlinear medium, other peaks in the transmissivity
occur when the nonlinearity induces a resonance.  We will refer to
them as {\em nonlinear modes}, i.e., nonlinear spatial patterns whose
fields oscillate everywhere at the same frequency inside the cavity. 
In this Letter we report the observation of a nonlinear mode in a
cylindrical Fabry-Perot cavity~\cite{deutsch1992} filled with atomic
rubidium vapor.  The cavity-field buildup and resonantly-enhanced
nonlinearity give stronger nonlinear effects for a given incident beam
intensity than do the traveling-wave configurations, permitting our
use of a low-power CW diode laser.

\begin{figure}
\centerline{\psfig{figure=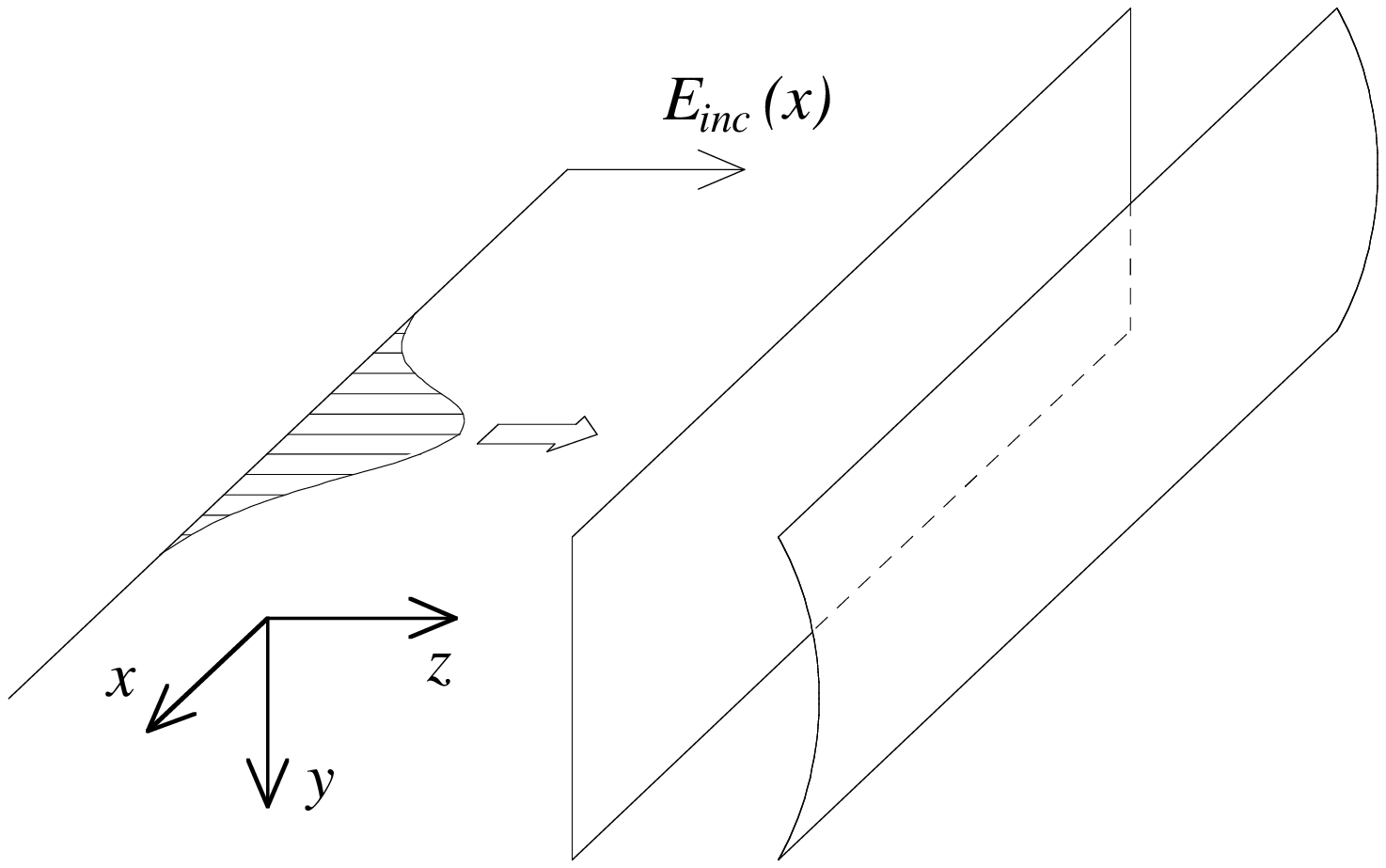,width=12cm}}
\caption{}
\label{cavityfigure}
\end{figure}

To study 1D transverse nonlinear optical effects, the experiment 
consists of a cylindrical Fabry-Perot cavity illuminated with an 
incident beam $E_{inc}$, as shown in Fig.  \ref{cavityfigure}, and 
filled with a material having a Kerr nonlinearity ($n = n_0+n_2 
|E|^2$).  The mirror curvature in the $y$-dimension causes strong 
single-moding in that direction, and a single longitudinal mode is 
excited.  The most important dynamics are in the $x$-dimension, where 
the field is determined by the combined effects of self-(de)focusing 
and diffraction.  The field envelope within the cavity, in the mean 
field limit, evolves according to a damped, driven version of the 
nonlinear Schr\"{o}dinger equation \cite{lugiato1987,boyce1999} 
{\small
\begin{equation}
\frac{\partial E}{\partial t} = 
\frac{i c}{2 n_0 k}\frac{\partial^2 E}{\partial x^2}
+ i\omega A\frac{n_2}{n_0} |E|^2 E \\
+ \frac{i c \Delta k}{n_0} E - \Gamma( E- E_d)\,,
\label{cavityeqn}
\end{equation}
} where $E(x,t)$ is the internal cavity field envelope amplitude, $k$ 
is the longitudinal wavenumber, $\omega=ck/n_0$ is the field angular 
frequency, $A=3/4\sqrt{2}$ is a mode-overlap factor, $\Delta k = k - 
k_{L}$ is the wavenumber mismatch within the cavity of a weak driving 
field from linear-response plane-wave resonance, $\Gamma=c({\cal 
T}+\alpha L)/2 n_{0}L$ is the amplitude decay rate (${\cal T}$ is the 
intensity transmission coefficient at each mirror, assumed equal, $L$ 
is the cavity length, $\alpha$ is the absorption coefficient due to 
the intracavity material, and ${\cal T}+\alpha L \ll 1$), and 
$E_{d}(x)$ is the driving field, which is proportional to the incident 
field.  In our experiment each of the terms on the right of Eq.  
\ref{cavityeqn} is comparable in magnitude.

\begin{figure}
\centerline{\psfig{figure=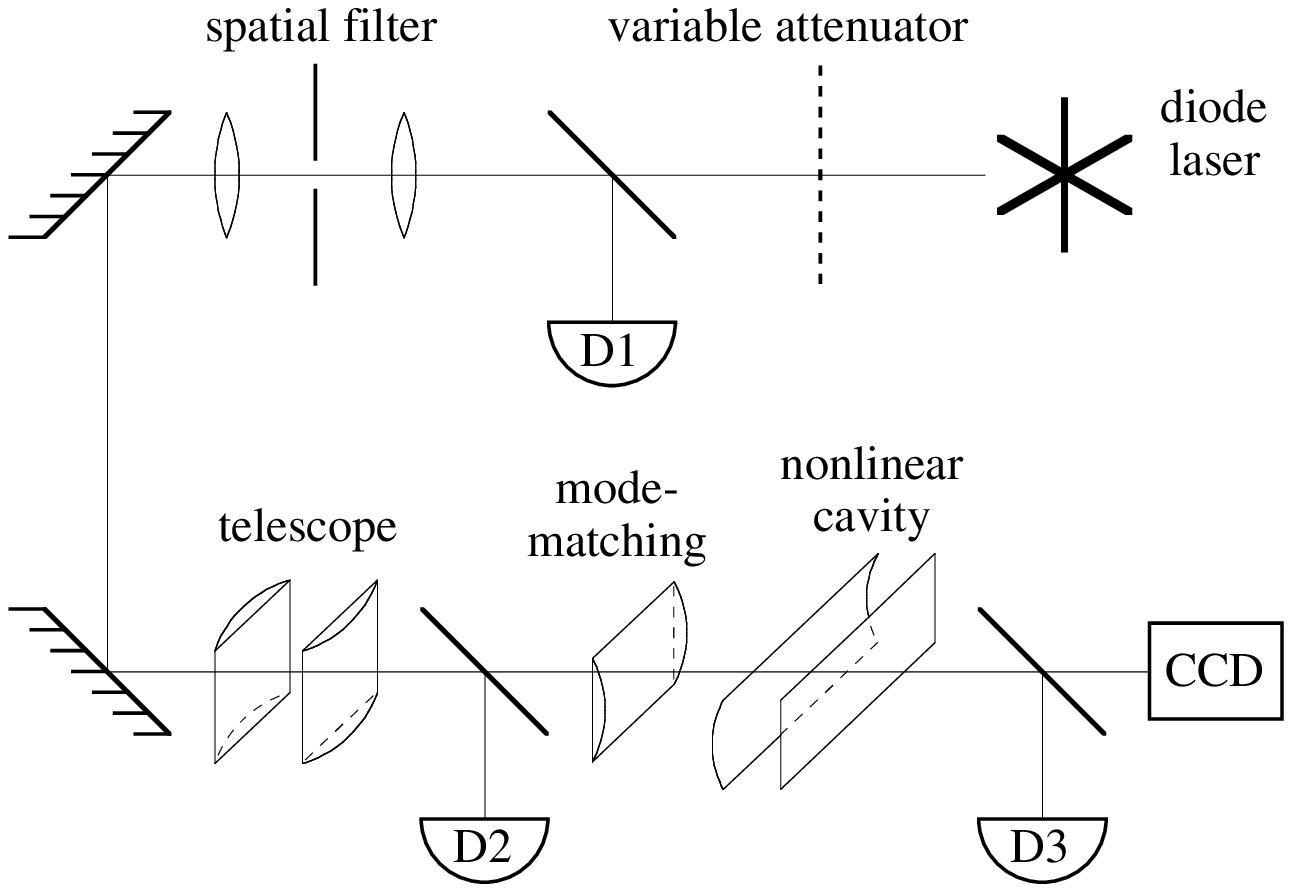,width=12cm}}
\caption{}
\label{exptfigure}
\end{figure}

The experimental arrangement used to observe the spatial nonlinear 
mode in a Fabry-Perot cavity is shown in Fig.  \ref{exptfigure}.  A 
$35\,\mathrm mW$, $780\,\mathrm nm$ Sharp diode laser with external 
grating feedback (PZT-tuned) is used as the light source.  After a 
variable attenuator and a spatial filter, the beam travels through a 
3-element cylindrical telescope used to control the beam waist in the 
$x$-direction to between 30 and 300 $\mu\textrm m$ at the cavity input 
face.  A cylindrical mode-matching lens focuses in the $y$-direction 
to efficiently couple into the cavity's lowest-order transverse mode.  
The CCD camera and photodiode D3 image the light leaving the exit 
mirror of the cavity.  Two additional detectors are also used in the 
experiment: detector D1 is used primarily for alignment (it only 
receives appreciable power when the beam is collimated in $x$ and 
mode-matched in $y$, at the cavity input face), and detector D2 
monitors input power.

\begin{figure}
\centerline{\psfig{figure=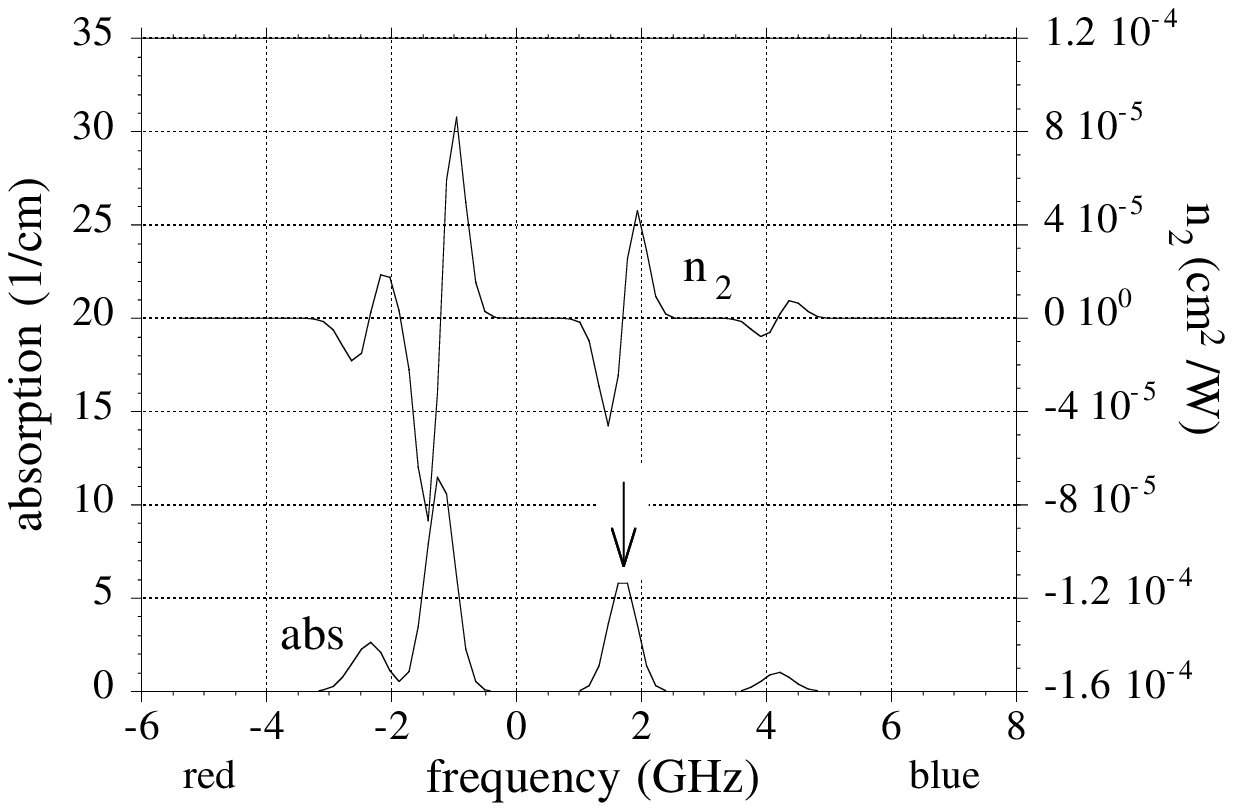,width=12cm}}
\caption{}
\label{n2figure}
\end{figure}

The nonlinear Fabry-Perot cavity itself consists of two mirrors, one 
cylindrical and the other planar, separated by $L=4.5\,\mathrm mm$ and 
with ${\cal T}\approx0.007$.  The cavity length and alignment are 
precisely tunable using 3 micrometers and piezoelectric elements on 
the planar mirror.  The concave mirror has a cylindrical radius of 
curvature of $9.8\,\mathrm mm$, yielding a mode beam waist (1/e 
half-width in field profile) of $35\,\mu\mathrm m$ in the 
$y$-direction.  The cavity is filled with natural-abundance rubidium 
vapor at an atomic number density of $N=1.5\times10^{12}\,{\mathrm 
cm}^{-3}$.  The laser light excites the atoms on their D2 transition 
near $780\,\mathrm nm$ to provide a resonantly-enhanced optical 
nonlinearity.  Figure \ref{n2figure} shows the calculated linear 
absorption and nonlinear index $n_2$, including Doppler broadening and 
all hyperfine magnetic sublevels.  The arrow in Fig.  \ref{n2figure} 
indicates the line used in the experiment, corresponding to 
transitions out of the $^{85}\mathrm Rb$, $F=2$ ground-state hyperfine 
level.  This line was chosen due to its relative strength and 
separation in frequency from other lines.

\begin{figure}
\centerline{\psfig{figure=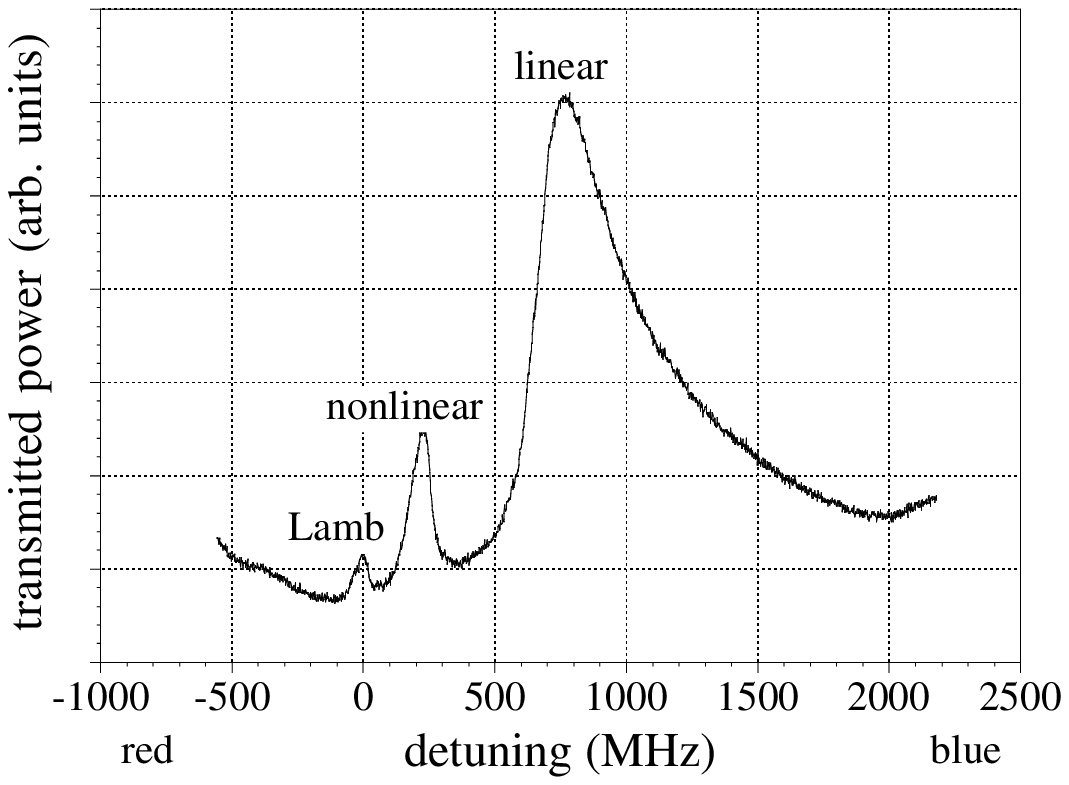,width=12cm}}
\caption{}
\label{scanfigure}
\end{figure}

Figure~\ref{scanfigure} shows an example of the variation in optical
power transmitted through the cavity as the laser frequency is scanned
at a rate of $20\,\mathrm{Hz}$, for $0.82\,\mathrm{mW}$ of incident
power (beam waists of $130\,\mu\textrm m$ and $35\,\mu\textrm m$ in
the $x$ and $y$ dimensions, respectively) and a fixed cavity length
$L$.  Bistability (a dependence on scanning direction) is not
observed.  The transmission peak marked ``linear'' corresponds to the
ordinary linear Fabry-Perot transmission resonance.  The asymmetry of
this resonance is primarily due to the linear index and absorption of
the atomic vapor.  The ``nonlinear'' peak is observed to be
power-dependent in two ways: firstly, it shifts in frequency as a
function of the incident laser intensity, and secondly, it disappears
at sufficiently low intensity.  This peak occurs where the nonlinear
index change is sufficient to shift the red-detuned laser into cavity
resonance.  As incident beam power is decreased, it moves toward
atomic line center, away from the linear resonance.  Once a maximum
frequency shift is reached, the nonlinear resonance peak smoothly
vanishes as the incident power is further decreased.  This peak is
thus the nonlinear mode.  By contrast, the linear mode is observed not
to change at all with power.  The ``Lamb'' feature is due to the
forward- and backward-going cavity beams saturating the atomic
absorption at line center.

\begin{figure}
\centerline{\psfig{figure=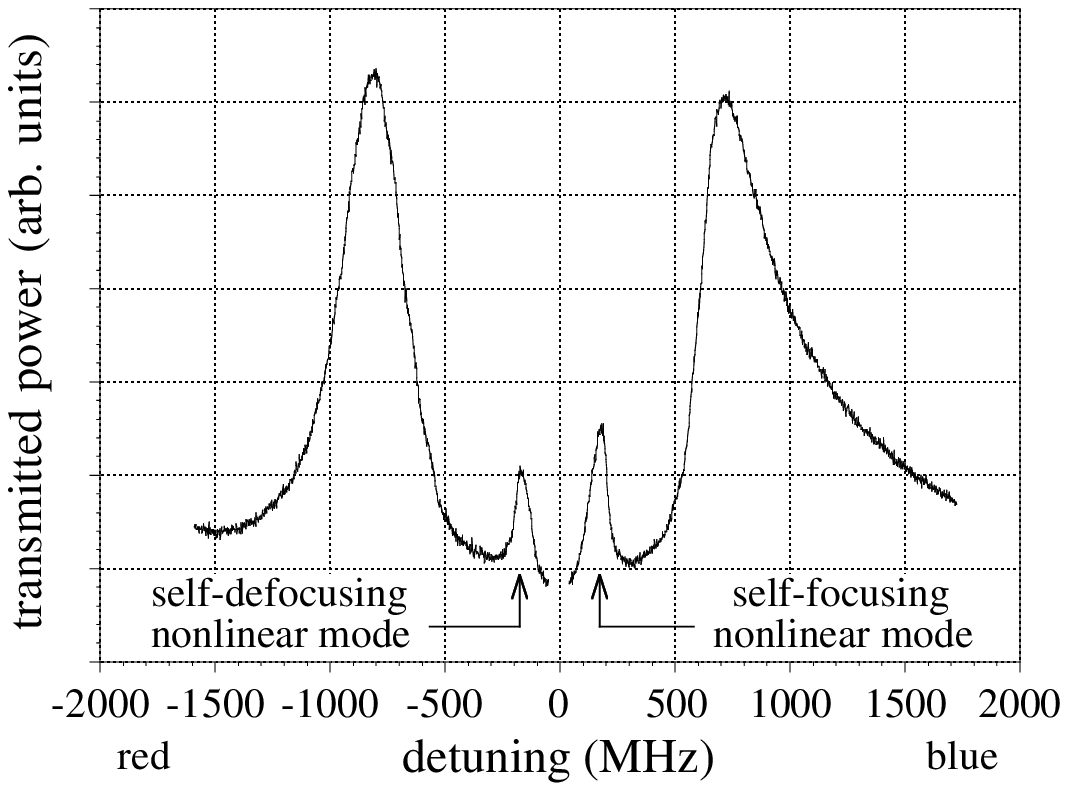,width=12cm}}
\caption{}
\label{asymmetryfigure}
\end{figure}

Transmission spectra for two different values of the cavity length $L$ 
are compared in Fig.  \ref{asymmetryfigure}, using the same incident 
beam power and dimensions as above.  There is a transmission 
enhancement in the nonlinear resonance on the blue side of the atomic 
line; this is due to the presence of self-focusing in the 
$x$-direction on that side, and self-defocusing on the other.  In the 
former case the effect acts against diffraction to give more complete 
constructive interference in the forward direction, whereas in the 
latter case the nonlinearity acts \textit{with} diffraction and the 
transmission is lower.  Absorption complicates the spatial aspects of 
nonlinear mode formation, but a comparison between the red- and 
blue-detuned cases in Fig.  \ref{asymmetryfigure} is meaningful since 
absorption is symmetrical about line center.

\begin{figure}
\centerline{\psfig{figure=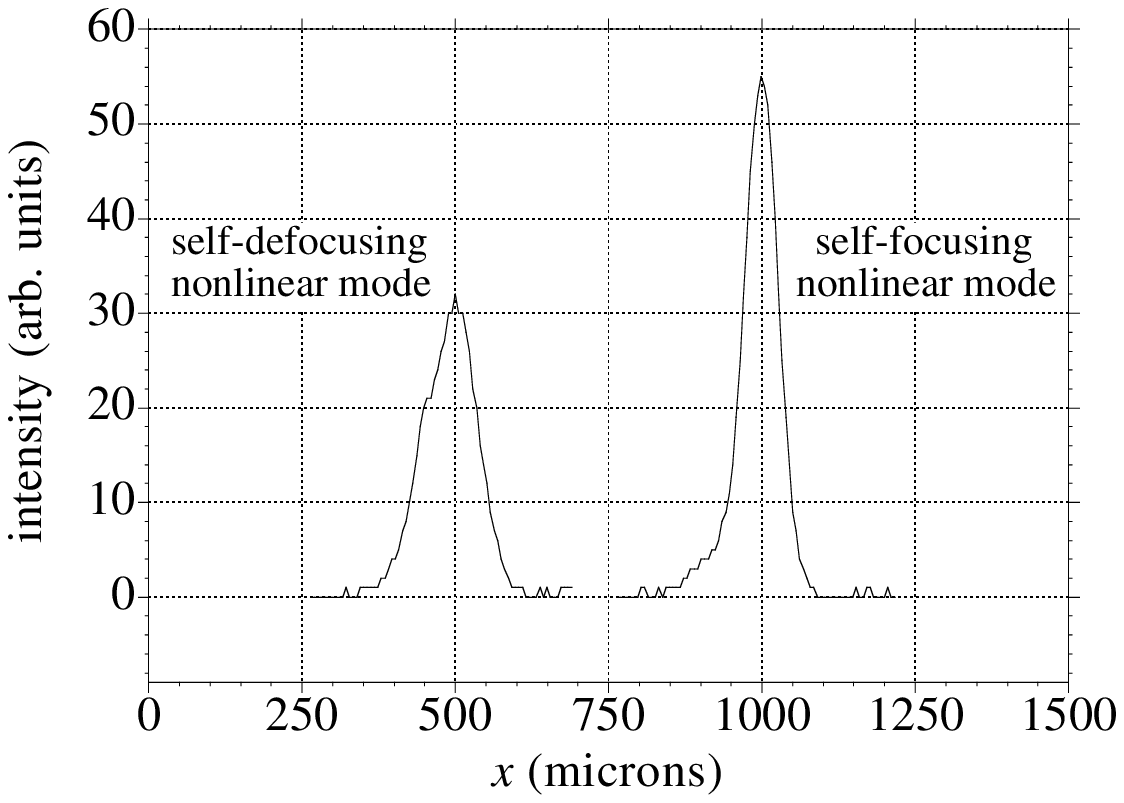,width=12cm}}
\caption{}
\label{profilesfigure}
\end{figure}

Spatial profiles of the transmitted beams for two cases with equal 
atomic absorptions and opposite signs of $n_{2}$ are shown in 
Fig.~\ref{profilesfigure}, again using $0.82\,\mathrm{mW}$ of incident 
power and Gaussian beam waists of $130\,\mu\textrm m$ and 
$35\,\mu\textrm m$ in the $x$ and $y$ dimensions, respectively.  As 
discussed above the transmission is substantially higher when 
$n_{2}>0$ and the width is also narrower in this case.  The wavenumber 
mismatch $\Delta k$ is measured using $c\Delta 
k=n_{0}(\omega_{NL})\omega_{NL}- n_{0}(\omega_{L})\omega_{L}$, where 
$\omega_{NL}$ and $\omega_{L}$ are the measured frequencies of the 
nonlinear and linear resonances, respectively.  We infer 
$n_{0}(\omega)$ by measuring the frequency splitting of a linear 
cavity mode tuned exactly on atomic resonance (in our case the mode 
divides into two components at detunings of roughly $\pm 
500\,\mathrm{MHz}$ relative to line center), and extrapolating to 
other frequencies using the known functional form of $n_{0}(\omega)$ 
in a Doppler-broadened atomic vapor.  The measured nonlinear mode 
widths $\sigma$ (defined as the root-mean-square widths of the 
$y$-integrated intensity profiles) for the range $10\,{\mathrm 
m}^{-1}<|\Delta k|<20\,{\mathrm m}^{-1}$ are $34.7\,\mu{\mathrm m} < 
\sigma < 39.2\,\mu{\mathrm m}$ for $n_{2}>0$ and $41.7\,\mu{\mathrm m} 
< \sigma < 44.9\,\mu{\mathrm m}$ for $n_{2}<0$, with the smaller 
widths corresponding to larger values of $|\Delta k|$.

In conclusion, we have demonstrated a new type of spatial nonlinear 
mode within a cylindrical nonlinear Fabry-Perot resonator.  In the 
self-focusing case, some features of this mode are similar to spatial 
solitons in traveling-wave systems -- numerical work indicates that 
the nonlinear mode has structural stability in the sense that the 
transmitted beam profile is insensitive to modest changes in the 
incident beam profile.  The relationship between the cavity nonlinear 
mode and spatial solitons would be in interesting topic for further 
research.

This work was done with the support of the Office of Naval Research 
under ONR Grant Number N00014-99-1-0096, and of the National Science 
Foundation under NSF Grant Number 9722535.  J.P. Torres is grateful to 
the Spanish Government for its support funded through the Secretar\'{\i}a 
de Estado de Universidades, Investigaci\'{o}n y Desarrollo.  We would also 
like to thank J. Bowie, J. Garrison, J. McGuire, and M. W. Mitchell 
for very helpful discussions.

\newpage
	
\newpage

\newpage

\newpage

\section*{FIGURE CAPTIONS}

\noindent
FIG 1.  The cylindrical Fabry-Perot cavity, showing the coordinate
system used and an incident Gaussian (in $x$) beam.
\vskip 1cm

\noindent
FIG 2.  Schematic of the experimental apparatus used to observe the spatial
nonlinear mode.  The D's are detectors.
\vskip 1cm

\noindent
FIG 3.  Calculated linear absorption and $n_2$ in natural-abundance Rb 
vapor at $80^{\circ}\mathrm{C}$ ($N=1.5\times 10^{12}\,{\mathrm 
cm}^{-3}$), near the D2 transition at $\lambda=780\,\mathrm nm$, for 
circularly-polarized light.  The arrow indicates the line used in the 
experiment, the Doppler-broadened $^{85}\mathrm Rb$, $F=2$ set of 
transitions.
\vskip 1cm

\noindent
FIG 4.  Power transmitted through the nonlinear cavity as a function 
of laser frequency, for a particular cavity length $L$ and beam 
power.  Frequency is relative to the $^{85}\mathrm Rb$ line indicated in
Fig. \ref{n2figure}.
\vskip 1cm

\noindent
FIG 5.  Total power transmitted through the cavity as a function of laser
frequency, for two different cavity lengths $L$.
\vskip 1cm

\noindent
FIG 6.  Transmitted intensity profiles of spatial nonlinear mode 
taken with the CCD camera, integrated over $y$, at the peaks of the 
nonlinear modes shown in Fig.~\ref{asymmetryfigure}.
\vskip 1cm

\end{document}